\RequirePackage{mathtools}
\documentclass{llncs}

\usepackage{placeins}
\usepackage[title]{appendix}

\usepackage{longtable}% for long tables

\usepackage[utf8]{inputenc}
\usepackage{amsfonts}
\usepackage{amssymb}
\usepackage{xspace}
\usepackage{paralist}
\usepackage{cleveref}
\crefname{algocf}{alg.}{algs.}
\Crefname{algocf}{Algorithm}{Algorithms}

\usepackage[commentColor=blue,rightComments=false]{algpseudocodex}
\usepackage{algorithm2e}

\usepackage{footmisc}
\usepackage{color}
\usepackage{tikz}
\usetikzlibrary{arrows, automata, positioning, fit, shapes}

\tikzset{%
  sh2n/.style={shift={(0,0.31)}},
  sh2s/.style={shift={(0,-0.31)}},
  sh2e/.style={shift={(0.31,0)}},
  sh2w/.style={shift={(-0.31,0)}},
  sh2nw/.style={shift={(-0.31,0.31)}},
  sh2ne/.style={shift={(0.31,0.31)}},
  sh2sw/.style={shift={(-0.31,-0.31)}},
  sh2se/.style={shift={(0.31,-0.31)}},
  rc/.style={rounded corners=2mm},
  reflexive above/.style={->,loop,looseness=7,in=60,out=120, above},
  reflexive below/.style={->,loop,looseness=7,in=240,out=300, below},
  reflexive left/.style={->,loop,looseness=7,in=150,out=210, left},
  reflexive right/.style={->,loop,looseness=7,out=30,in=330, right},
}

\graphicspath{{figures/}}

\SetStartEndCondition{ }{}{}%
\SetKwFunction{Range}{range}%%
\SetKw{KwTo}{in}\SetKwFor{For}{for}{\string:}{}%
\SetKwIF{If}{ElseIf}{Else}{if}{ then}{elif}{else}{}%
\SetKwFor{While}{while}{:}{fintq}%
\SetKwRepeat{Repeat}{\addtocounter{AlgoLine}{-1}repeat}{\nl until}%
\SetKwFor{For}{for}{ do}{}%
\SetKw{Return}{return}%

\AlgoDontDisplayBlockMarkers\SetAlgoNoEnd%\SetAlgoNoLine%
\DontPrintSemicolon

\SetKwFunction{Backbone}{\textrm{\textsc{Backbone}}}
\SetKwFunction{Home}{\textrm{\textsc{Home}}}
\SetKwProg{Fn}{Function}{}{}
\SetKw{Init}{Initializing:}

\usepackage{graphics}

\usepackage[draft,layout=inline]{fixme}
\FXRegisterAuthor{rg}{arg}{RG}
\FXRegisterAuthor{as}{aas}{AS}
\FXRegisterAuthor{mf}{amf}{MF}
\FXRegisterAuthor{nw}{anw}{NW}

\usepackage{todonotes}

\newcommand{\hW}{\textit{hW}-inference\xspace}
\newcommand{\ehW}{\textit{ehW}-inference\xspace}
\newcommand{\hND}{\textit{h}-ND\xspace}
\newcommand{\WND}{\textit{W}-ND\xspace}
\newcommand{\splitatcommas}[1]{%
  \begingroup
  \begingroup\lccode`~=`, \lowercase{\endgroup
    \edef~{\mathchar\the\mathcode`, \penalty0 \noexpand\hspace{0pt plus 1em}}%
  }\mathcode`,="8000 #1%
  \endgroup
}

\newcommand{\negativespace}{\vspace*{-6pt}}

\usepackage{subfig}
\begin{document}

\title{Active Inference of Extended Finite State Machine Models with Registers and Guards}

\author{
    Roland Groz \inst{1} \and
    Germán Vega \inst{1} \and
    Adenilso Simao \inst{3} \and
    Catherine Oriat  \inst{1} \and
    Neil Walkinshaw \inst{2} \and
    Michael Foster \inst{2}
    }

\institute{LIG, Université Grenoble Alpes, F-38058 Grenoble, France,
\email{Roland.Groz@univ-grenoble-alpes.fr},
\email{Catherine.Oriat@univ-grenoble-alpes.fr}
\email{german.vega@imag.fr},
\and Department of Computer Science, The University of Sheffield, UK
\email{n.walkinshaw@sheffield.ac.uk},
\email{m.foster@sheffield.ac.uk}
\and Universidade de S\~{a}o Paulo, ICMC, S\~{a}o Carlos/S\~{a}o
 Paulo, Brasil \email{Adenilso@icmc.usp.br}}

\maketitle

\begin{abstract}

Extended finite state machines (EFSMs) model stateful systems with internal data variables and have numerous applications in software engineering. A major advantage of this type of model lies in its ability to model both the data flow and the data-dependent control behaviour. In the absence of such models, it is desirable to reverse-engineer them by observing the system's behaviour. However, existing approaches generally require the ability to reset the system during inference, or can only handle situations where the control flow depends exclusively on the input parameters, and not on the values of the stored data. In this work, we present a black-box active learning algorithm that infers EFSMs with guards and registers, and which significantly relaxes the assumptions that have to be made about the system in comparison to previous attempts.

\end{abstract}

\section{Introduction}\label{sec:introduction}
Accurate models of software behaviour are useful for a wide range of software engineering tasks, such as validating system correctness \cite{groce2002adaptive}, generating test inputs \cite{choi2013guided}, and comparing behaviour between different software versions \cite{damasceno2019learning}.
Reactive systems --- systems that respond to their environment, their users, or other systems --- are often modelled as (Extended) Finite State Machines ((E)FSMs), and such models form the basis of many testing and verification techniques \cite{lee1996principles}.
EFSMs provide a semantic model that is at the heart of behavioural modelling of systems in such formalisms as SDL, StateCharts, and UML state diagrams, that can be regarded as various syntaxes for the similar concepts.

Despite their value, there are many software development contexts where models are overlooked or are not updated after an initial design phase.
In this case, it can be useful to \emph{reverse engineer} a model from the system itself. The task of inferring (E)FSM models has been the subject of a considerable amount of research. A popular strategy is the Minimally Adequate Teacher framework \cite{angluin1988queries}, where a model is inferred by posing a series of \emph{queries} to the system under inference (aka SUL, System Under Learning).

However, existing inference techniques \cite{foster2022reverse,howar2012inferring,isberner2014learning,walkinshaw2016inferring} tend to pose these queries from some known ``initial'' state, implicitly requiring the ability to reset the SUL. However, resetting the system is not always feasible, and is often problematic or costly.
Several inference approaches have been developed to avoid resets \cite{rivest1989inference,groz2020hw}.
Nonetheless, these approaches do not consider how data values affect the behaviour of the SUL, as they are restricted to finite automata or FSM.

In previous work \cite{Foster2023ICFEM}, we have proposed a two-stage approach to tackle this challenge:
an extension of the hW algorithm \cite{groz2020hw} is used to infer the control structure without reset, and genetic programming \cite{DBLP:books/lulu/PoliLM2008} is used to infer guards, register updates, and output functions from the corresponding data observations. That previous approach however did not support the use of registers in transition guards. In this paper, we overcome that limitation.

The main contributions of this paper are as follows:
\begin{compactitem}
    \item An algorithm that does not require to reset the system, learning from a single trace by sending inputs and observing outputs; it is able to reverse engineering EFSM models of software systems in full black box mode, with no required knowledge of the system apart from its interface.
%An extension to the \ehW algorithm \cite{Foster2023ICFEM} which supports registers in transition guards.
 %   \item A ``proof of concept'' demonstration of the algorithm being applied to a small example system.
    \item Relatively weak assumptions about the system (most notably, observability, and some level of boundedness when exercising the system with a small subset of inputs), when compared with previous work.
    \item No restriction on data types or the types of functions used for computing output parameters or evaluating guards of transitions.
\end{compactitem}

The rest of the paper is structured as follows.
\Cref{sec:background} and \ref{sec:definitions} presents a motivating example and a brief formalisation of EFSMs.
\Cref{sec:assumptions} lays out the main assumptions our algorithm makes about the SUL.
\Cref{sec:algorithm} presents an overview of our inference algorithm.
\Cref{sec:example} provides a walk-through of the algorithm, showing how it can be applied to our motivating example from \Cref{sec:background}.
Finally, \Cref{sec:conclusion} concludes the paper and discusses potential future work.

\section{Running example}\label{sec:background}

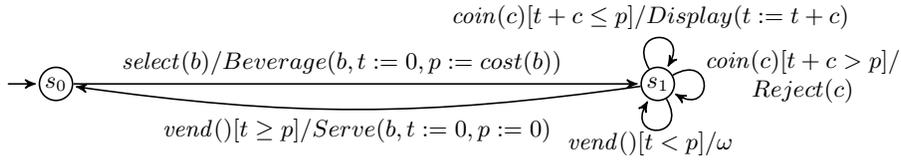
\begin{figure}[!ht]{%
    \subfloat{%
      
%      \centering
      \hspace*{-0.4em}\begin{tikzpicture}[->,>=stealth',shorten >=1pt,auto, node distance=5cm, semithick, initial left, initial text={},
    state/.style={circle, draw, minimum size=0.3cm, inner sep=1}]
    \node[initial, state] (s0)               {$s_0$};
    \node[state, shift={(3, 0)}]          (s1) [right of=s0] {$s_1$};

    \draw (s0) edge [above]node[align=center, shift={(-0.2, 0.0)}]{$\mathit{select}(b)/Beverage(b,t:=0,p:=cost(b))$}  (s1);
    \draw (s1) edge [reflexive above, above]node[align=center, shift={(-0.1, 0.0)}]{$\mathit{coin}(c)[t+c \le p]/Display(t:=t+c)$} (s1);
    \draw (s1) edge [reflexive right]node[align=center, shift={(-0.1, 0.1)}]{$\mathit{coin}(c)[t+c > p]/$\\$Reject(c)$} (s1);
    \draw (s1) edge [reflexive below]node[align=center, shift={(-0.1, 0.1)}]{$\mathit{vend}()[t < p]/\omega$} (s1);
    \draw (s1) edge [bend left=8, below]node{$\mathit{vend}()[t \ge p]/Serve(b,t:=0,p:=0)$}  (s0);
\end{tikzpicture}
%      \caption{Running example: registers used in guards}
    }
  }
  \caption{Our vending machine EFSM and an example trace.}% caption for whole figure
  \negativespace
  \label{fig:drinks}% label for whole figure
  \label{fig:runningexample}
\end{figure}

We consider a straightforward running example that exhibits most features of EFSMs, yet is small enough to be fully illustrated in this paper. Our example is a vending machine, shown in Fig.~\ref{fig:drinks}. 

Parametrised input events represent interactions initiated by the user or the environment; for instance, \textit{select(b)} represents the choice of a drink \textit{b} and \textit{coin(c)} the insertion of a coin with value \textit{c}. Similarly, parametrised output events represent the response of the system; for instance, \textit{Beverage(b,t,p)} is the response to the selection of a drink \textit{b} and shows its price \textit{p} and \textit{Display(t)} shows the total amount of money \textit{t} currently paid after a coin is inserted. When an input type is not accepted in a state (such as \textit{coin} or \textit{vend} in $s_0$), there is an implicit loop transition with that input and output $\Omega$. When the input is not refused in a state, the output $\omega$ indicates that the system produces no visible output nor changes the system state.

When querying the system we observe input/output event \emph{occurrences} with valued arguments for the parameters, Fig.~\ref{fig:example-trace} shows an example of an execution trace. Throughout the paper, we refer to the valued input event occurrences --- such as $select(tea)$ --- as concrete inputs of the system and to the corresponding event types --- such as $select$ --- as abstract inputs, similarly for output events.

\begin{figure}[!hb]{%

    \subfloat{%
      
      \centering
      \parbox{\textwidth}{\centering $ select(tea)/Beverage(tea,0,100), coin(50)/Display(50), vend()/\omega,$ \newline
        $ coin(50)/Display(100),$
        $coin(50)/Reject(50), vend()/Serve(tea,0,0) $}
    }
  }
  \caption{Example of an execution trace for the running example.}% caption for whole figure
  \negativespace
  \label{fig:example-trace}
\end{figure}
For every input/output parameter there is a corresponding register that stores the last observed value of the parameter. By convention, register are named with the same name of the corresponding parameter (there are then four registers in the drinks machine: \textit{b}, \textit{p}, \textit{c} and \textit{t}) and are automatically updated from the values of input/output event arguments. Registers are used in \emph{guards} and \emph{output expressions}, and can then influence later computations. 

We call this particular style of EFSM an input/output register machine, and this is the kind of machines that our algorithm is able to infer, see the discussion on \emph{register observability} in \Cref{sec:assumptions}.

We remark that this machine could not be inferred by the method in \cite{Foster2023ICFEM}, because the response of the system to the insertion of a coin or to pressing the $vend$ button depends on the recorded value of the total amount paid into the machine, which can only be described by a register.

\section{Definitions}\label{sec:definitions}
In this work, we formally define an EFSM as a tuple $(Q,\mathcal{R},\mathcal{I},\mathcal{O},\mathcal{T})$ where
$Q$ is a finite set of states,
$\mathcal{R}$ is a cartesian product of domains, representing the types of registers. There is a finite number of registers, but their domains can be infinite.
$\mathcal{I}$ is the set of concrete inputs, structured as a finite set of abstract inputs $I$ each having associated parameters and their domains $P_I$. Similarly $\mathcal{O}$ is the set of concrete outputs based on $O$ for abstract outputs.
$\mathcal{T}$ is a finite set of transitions, i.e. tuples
$(s,x,y,G,F,U,s')$ where
$s, s' \in Q$, $x \in I$, $y \in O$,
$G: P_I(x) \times \mathcal{R} \rightarrow \mathbb{B}$ is the transition guard,
$F: P_I(x) \times \mathcal{R} \rightarrow P_O(y)$ is the output function that gives the value of the output parameters,
$U: P_I(x) \times \mathcal{R} \rightarrow \mathcal{R}$ is the update function that gives the value of the registers after the transition.

\begin{figure}[!ht]{%
    \subfloat{%
      
     \centering
       \begin{tikzpicture}[->,>=stealth',shorten >=1pt,auto, node distance=5cm, semithick, initial left, initial text={},
    state/.style={circle, draw, minimum size=0.3cm, inner sep=1}]
    \node[initial, state] (s0)               {$s_0$};
    \node[state, shift={(3, 0)}]          (s1) [right of=s0] {$s_1$};

    \draw (s0) edge [above]node[align=center, shift={(-0.2, 0.0)}]{$\mathit{select}/Beverage$}  (s1);
    \draw (s1) edge [reflexive above, above]node[align=center, shift={(-0.1, 0.1)}]{$\mathit{coin}/Display$} (s1);
    \draw (s1) edge [reflexive below]node[align=center, shift={(-0.1, 0.1)}]{$\mathit{coin}/$\\$Reject$} (s1);
    \draw (s1) edge [reflexive right]node[align=center, shift={(-0.1, 0.1)}]{$\mathit{vend}/\omega$} (s1);
    \draw (s1) edge [bend left=8, below]node{$\mathit{vend}/Serve$}  (s0);
\end{tikzpicture}%
    }
  }
  \caption{Control NFSM of the vending machine.}
  \label{fig:vending}
  \label{fig:vendingNFSM}
\end{figure}
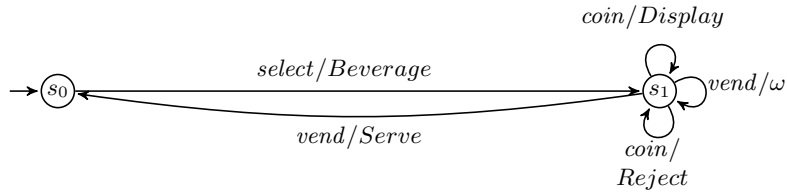

Associated with an EFSM, its \emph{control machine} is the non-deterministic finite state machine (NFSM) that results from abstracting from parameters (and therefore guards and registers). It defines the finite control structure of a given EFSM and is defined by a quadruple $(Q, I, O,\Delta)$, where $\Delta$ is the transition function
$\Delta: Q \times I \times O \to Q$. The control NFSM corresponding to our drinks vending machine EFSM is shown in Fig.~\ref{fig:vendingNFSM}.

\begin{figure}[!ht]{%
    \subfloat{%
      
      \centering
      \begin{tikzpicture}[->,>=stealth',shorten >=1pt,auto, node distance=5cm, semithick, initial left, initial text={},
    state/.style={circle, draw, minimum size=0.3cm, inner sep=1}]
    \node[initial, state] (s0)               {$s_0$};
    \node[state, shift={(2.5, 0)}]          (s1) [right of=s0] {$s_1$};

    \draw (s0) edge [above]node[align=center, shift={(-0.2, 0.0)}]{$\mathit{select}/Beverage$\\[-0.5mm]
    \begin{tabular}{|l|l||l|l|}
    \hline
     $(\bot,\bot,\bot,\bot)$&\textit{tea}&(\textit{tea},0,100)&$(\textit{tea},\bot,0,100)$\\
     \hline
     \end{tabular}
    }  (s1);
    \draw (s1) edge [reflexive above, above]node[align=center, shift={(-0.1, 0.1)}]{$\mathit{coin}/Display$\\
    \begin{tabular}{|l|l||l|l|}
    \hline
      (\textit{tea},$\bot$,0,100)&50&50 &(\_,50,50,\_)\\
      (\textit{tea},50,50,100)&50&100&(\_,50,100,\_)\\
    \hline
    \end{tabular}
    } (s1);
    \draw (s1) edge [reflexive below]node[align=center]{$\mathit{coin}/Reject$\\
    \begin{tabular}{|l|l||l|l|}
    \hline
      (\textit{tea},50,100,100)&50&50&(\_)\\
    \hline
    \end{tabular}
    }  (s1);
    \draw (s1) edge [reflexive right]node[align=center]{$\mathit{vend}/\omega$\\
    \begin{tabular}{|l||l|}
    \hline
    (\textit{tea},50,100,100)&(\_)\\
    \hline
    \end{tabular}
    } (s1);
    \draw (s1) edge [bend left=8, below]node[shift={(-0.4,0.0)}]{$\mathit{vend}/Serve$
    \begin{tabular}{|l||l|l|}
    \hline
    (\textit{tea},50,100,100)&\textit{tea}&(\_,\_,0,0)\\
    \hline
    \end{tabular}
    }  (s0);
\end{tikzpicture}%      \caption{Example of an execution trace for the running example}
    }
  }
  \caption{Sampled FSM with trace from Fig.~\ref{fig:example-trace}}% caption for whole figure
  \label{fig:vendingsampled}
\end{figure}
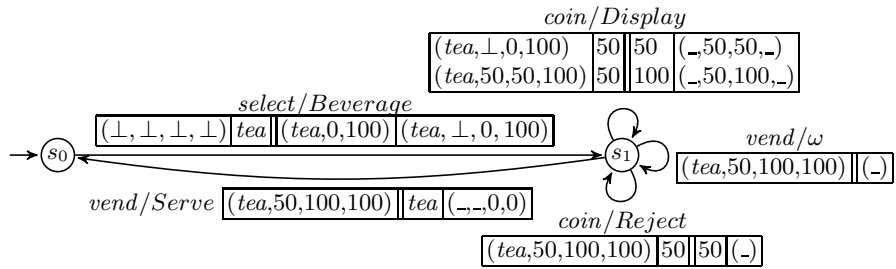

When we exercise a system, we learn \emph{occurrences} of transitions with concrete values for input and output parameters. For each transition of a control machine, we may collect several sample values. We call a \emph{sampled FSM} the quintuple $(Q,\mathcal{I},\mathcal{O},\Delta,\Lambda)$ where $\Lambda: Q \times I \times O \to 2^{\mathcal{R} \times \mathcal{I} \times \mathcal{O} \times \mathcal{R}}$ associates transitions from the control machine to a table of observed concrete inputs and outputs in a given register configuration.  The sampled FSM corresponding to our drinks vending machine EFSM and the trace of Fig.~\ref{fig:example-trace} is shown in Fig.~\ref{fig:vendingsampled}.

Our algorithm will first try to learn the control graph of the machine, then it will collect samples from observing inputs and outputs and record them in a sampled FSM. Finally, it uses all the samples available and genetic programming (symbolic regression) to infer the expressions and guards associated to transitions, to yield a conjecture EFSM, which should be behaviourally equivalent to the system.

\section{Assumptions}\label{sec:assumptions}

In order to guarantee the inference of a model that is exactly correct, and to ensure that the algorithm terminates in doing so, we need to make some assumptions about the nature of the system. We assume that the SUL is semantically equivalent to an EFSM, which has the following properties:
\begin{itemize}
  \item It is \emph{deterministic} (at the concrete level), and its control NFSM is \emph{strongly connected} (since we do not assume a reset, we can only learn a strongly connected component) and \emph{observable} in that if an input triggers different transitions from the same state, they have different abstract outputs.
  \item Registers can only take values from input and output parameters, and are therefore observable (\emph{register observability}): no hidden values influence the computation unnoticed. We further restrict them to storing only each parameter's last value (it could be extended to a bounded history). This enforces a particular style of machine modelling, in which update functions simply reflect the outputs, however the guards and output functions can be arbitrarily complex, so this does not restrict the expressive power. Note also that this assumption implies that, during learning, the set of potential registers is known, as it is given by the parametrised input/output alphabets.
  \item We also assume that the set of values taken by output parameters when the system is exercised with the restricted sample of concrete inputs is finite. \footnote{For instance, for a vending machine, we would not allow an unlimited amount of coins to be inserted. This is not a limitation in this example, since no physical machine could accept infinitely many coins.}
\end{itemize}

We also rely on Angluin's MAT (Minimally Adequate Teacher) inference framework~\cite{angluin1988queries}, i.e. we assume an oracle can provide a counterexample if there is a behavioural difference between the model and the system. This can be approximated by a random walk on the inferred machine.

\section{Overview of the algorithms}\label{sec:algorithm}

\Cref{alg:full-algo} (see \ref{app:algorithms}) outlines our \ehW{} algorithm. At its core, it adapts the Mealy \hW{} algorithm by \cite{groz2020hw}, based on a so-called \textsc{backbone}, to learn the control structure of the EFSM on abstract inputs and outputs. 

As the domains of input parameters can be infinite or very large (e.g. integers or floats), we will use sample values in the learning process. We assume we are given (or pick) two levels of samples of concrete inputs $I_1 \subset I_s \subset \mathcal{I}$, with $I_1$ used for learning the control structure and $I_s$ being a larger set of samples used to build a sampled FSM.  

The fact that registers can influence guards has two implications. First, from a given state, the same concrete input can produce two different abstract outputs. Second, and more importantly, characterisations (abstract responses to $W$) may be influenced by the register configuration. Therefore during the learning process, we can only conservatively infer that we are in the same state if we observe the same characterisation \emph{from the same register configuration}. 

Another difference is that to be able to incrementally characterise a state (or learn a transition from it) we need to reach it with the same register configuration. For this reason, we require a stronger notion of homing that always resets the registers, and there is additional complexity for transferring to the next state or transition to learn (as reflected in Algorithm \ref{func:transfer}).

Once we have found the control structure (NFSM), we walk the graph to try each input from $I_s$ in each transition, so as to collect enough samples for the \textsc{generalise} procedure to infer guards and output functions to build the EFSM. Generalisation into an EFSM (from the sampled FSM) proceeds as in \cite{Foster2023ICFEM}. The task here is simpler since we assume registers only store the last values of each parameter, so there is no need to infer the register update functions. 

The algorithm may learn redundant states since it can only assume that two inferred states are the same if they have been characterised from the same reference configuration of registers in $R$. However, in the worst case, this is still small compared to the automaton that comes from enumerating all possible configurations (with potentially an infinite state space). Once we have learnt all redundant copies of states fully, we can merge states that are are equivalent (in the NFSM structure) and compatible (in the sampled FSM): this is done by the \textsc{reduceFSM} procedure.

\subsection{Inconsistencies and counter-examples}
\label{sec:inconsistencies}

The algorithm incrementally builds the sampled FSM based on some hypothesized $h$ and $W$, however these tentative values may turnout to be wrong. This case can be detected by the algorithm by observing inconsistencies between the output of the SUL and the partially inferred FSM. Inconsistencies can be evidenced either during the learning phase of the algorithm while querying the SUL, or after inference by an oracle providing a counter-example. Depending on the nature of the inconsistency, the algorithm may refine $h$ (we call this an $h$-ND inconsistency), refine $W$ (a $W$-ND inconsistency), or simply discover a new transition that was not previously observed. For a full account of inconsistency handling see \cite{groz2020hw}.

\section{Execution of method on running example}\label{sec:example}

\newcommand{\event}[2]{%
	\begin{array}[b]{c}
		{\scriptstyle #1}\\ [-1ex]
	   \rightarrow {_{\scriptstyle #2}}
	\end{array}
}

\newcommand{\Q}[2]{%
	#2,\register{#1}
}

\newcommand{\register}[1]{%
	\langle#1\rangle
}

\newcommand{\state}[2][?]{%
	\begin{array}[c]{c}
		{\scriptstyle #2}\\
		\hline {\scriptstyle #1}
	\end{array}
}

\newcommand{\step}[2]{\underbrace{#2}_{#1}}
\newcommand{\h}[2][h]{\step{#1}{#2}}
\newcommand{\W}[2][W]{\step{#1}{#2}}
\newcommand{\transfer}[2][transfer]{\step{#1}{#2}}
\newcommand{\X}[2][X]{\step{#1}{#2}}
\newcommand{\s}[2][s]{\step{#1}{#2}}
\newcommand{\CE}[2][CE]{\step{#1}{#2}}

We assume we just know the interface of the system and have no clue about a homing sequence or a characterizing set, so we start with tentative $h=\epsilon$ and $W=\{\}$. We pick $I_1=\{\textit{coin}(100),\textit{select}(\textit{coffee}),\textit{vend}\}$ and we pick $I_s=I_1 \cup \{\textit{coin}(50),\textit{select}(\textit{tea})\}$. 

\begin{figure}[!ht]
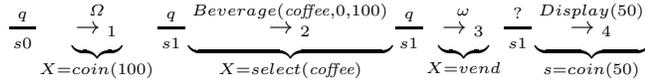

	\begin{tabular}[c]{l}
	$
\state[s0]{q}
\X[X=coin(100)]{
\event{\Omega}{1}
}
\state[s1]{q}
\X[X=select(\textit{coffee})]{
\event{Beverage(\textit{coffee},0,100)}{2}
}
\state[s1]{q}
\X[X=vend]{
\event{\omega}{3}
}
\state[s1]{?}
\step{s=coin(50)}{
\event{Display(50)}{4}
}
	$
	\end{tabular}
	\caption{Learning initial daisy, and first sample}
     \negativespace
 \label{fig:trace-a}
\end{figure}

Fig.~\ref{fig:trace-a} shows the initial steps of the algorithm. This sequence illustrates the overall behaviour of the algorithm : first start by homing to a known state, then try characterising the state reached, and from there try to learn the (abstract) output associated to each (concrete) input from $I_1$. 

At the end of the first iteration of the algorithm in Step~3, we end up with a first conjectured control structure : a ``daisy'' with a single state and all the transitions looping back. The algorithm starts now the sampling phase with $I_s$. The first sample, ($coin(50)$ in Step~4) yields output $Display$ instead of the expected abstract output $\Omega$ from the conjecture. This is a typical case of inconsistency. 

When processing this inconsistency, we realise that $coin$ separates at least two states of the control machine, so our initial guesses for $h$ and $W$ were necessarily  wrong, we refine our guess to $h=coin(100)$ and $W={coin(100)}$ and restart the backbone algorithm with these new values of $h$ and $W$ to try to identify a new conjectured control structure.

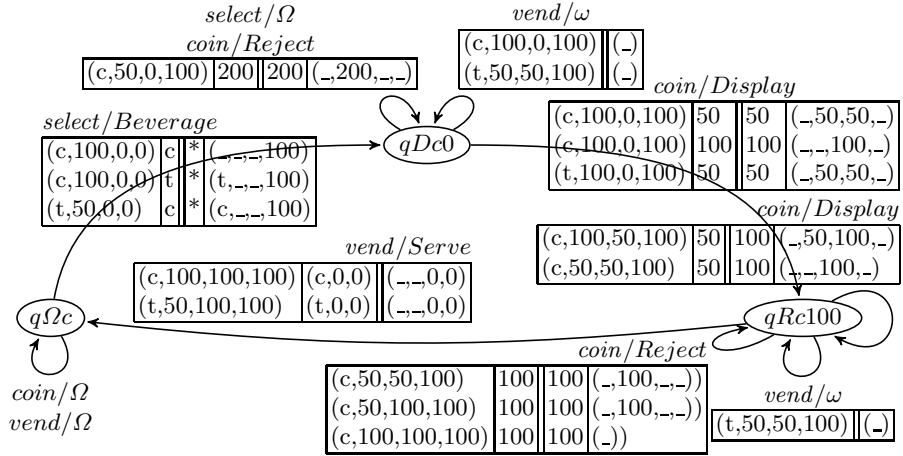
\begin{figure}[ht!]
    \centering
    \begin{tikzpicture}[->,>=stealth',shorten >=1pt,auto, node distance=5cm, semithick, initial left, initial text={},
    state/.style={circle, draw, minimum size=0.3cm, inner sep=1}]
    \node[state, shape=ellipse]                  (qOmc)                   {$q \Omega c$};
    \node[state, shape=ellipse, shift={(0, 2.3)}]  (qDc0)   [right of=qOmc] {$qDc0$};
    \node[state, shape=ellipse, shift={(0, -2.3)}] (qRc100) [right of=qDc0] {$qRc100$};

    \draw (qOmc) edge [reflexive below]node[align=center, shift={(0.0,0)}]{$coin/\Omega$\\$vend/\Omega$}  (qOmc);
    \draw (qOmc) edge [out=80, in=180]node[align=left, shift={(2, -2.7em)}]{
      $\mathit{select}/Beverage$\\
      \begin{tabular}{|l|l||l|l|}
      \hline
      (c,100,0,0)&c&*&(\_,\_,\_,100)\\
      (c,100,0,0)&t&*&(t,\_,\_,100)\\
      (t,50,0,0) &c&*&(c,\_,\_,100)\\
      \hline
      \end{tabular}
    }  (qDc0);
    \draw (qDc0) edge [reflexive above, out=150, in=110]node[align=center, shift={(-1.8, 0.0)}]
    {$\mathit{select}/\Omega$\\$\mathit{coin}/Reject$\\
    \begin{tabular}{|l|l||l|l|}
      \hline
      (c,50,0,100)&200&200&(\_,200,\_,\_)\\
      \hline
    \end{tabular}} (qDc0);
    \draw (qDc0) edge [reflexive above, out=30, in=70]node[align=center, shift={(1.1, 0)}]
    {$\mathit{vend}/ \omega$\\
    \begin{tabular}{|l||l|}
      \hline
      (c,100,0,100)&(\_)\\
      (t,50,50,100)&(\_)\\
      \hline
      \end{tabular}
    } (qDc0);
    \draw (qDc0) edge [above, out=0, in=100]node[align=center, shift={(0.6, -0.4)}]{
    $\mathit{coin}/Display$\\
    \begin{tabular}{|l|l||l|l|}
      \hline
      (c,100,0,100)&50 &50 &(\_,50,50,\_)\\
      (c,100,0,100)&100&100&(\_,\_,100,\_)\\
      (t,100,0,100)&50 &50 &(\_,50,50,\_)\\
      \hline
    \end{tabular}
    } (qRc100);
    \draw (qRc100) edge [reflexive below, out=210, in=190, left]node[align=right, shift={(0, -2em)}]{$\mathit{coin}/Reject$\\
    \begin{tabular}{|l|l||l|l|}
      \hline
      (c,50,50,100)  &100&100&(\_,100,\_,\_))\\
      (c,50,100,100) &100&100&(\_,100,\_,\_))\\
      (c,100,100,100)&100&100&(\_))\\
      \hline
    \end{tabular}
     } (qRc100);
    \draw (qRc100) edge [reflexive right, above]node[align=right, shift={(-2.3,1em)}]{$\mathit{coin}/Display$\\
    \begin{tabular}{|l|l||l|l|}
      \hline
      (c,100,50,100)&50&100&(\_,50,100,\_)\\
      (c,50,50,100) &50&100&(\_,\_,100,\_)\\
      \hline
    \end{tabular}
    } (qRc100);
    \draw (qRc100) edge [reflexive below]node[align=center]{$\mathit{vend}/ \omega$\\
    \begin{tabular}{|l||l|}
      \hline
      (t,50,50,100)&(\_)\\
      \hline
    \end{tabular}
    } (qRc100);
    \draw (qRc100) edge [above, bend left=6]node[align=right, shift={(-1.5, 0.15)}]{$\mathit{vend}/Serve$\\
    \begin{tabular}{|l|l||l|}
      \hline
      (c,100,100,100)&(c,0,0)&(\_,\_,0,0)\\
      (t,50,100,100) &(t,0,0)&(\_,\_,0,0)\\
      \hline
    \end{tabular}
    }  (qOmc);
\end{tikzpicture}
    \caption{The sampled FSM after 47 steps. }
  \negativespace
    \label{fig:drinks-nfsm}
\end{figure}

For lack of space we cannot present in detail the rest of the iterations of the algorithm (the full commented execution of the algorithm is presented in \ref{app:drinks_inference} for the interested reader), but after 47 steps, and with just two calls to the oracle to provide a counterexample, the final conjectured sampled FSM inferred by the algorithm is shown in Fig.~\ref{fig:drinks-nfsm}. 

\begin{figure}[ht!]
    \centering
    \begin{tikzpicture}[->,>=stealth',shorten >=1pt,auto, node distance=5cm, semithick, initial left, initial text={},
    state/.style={circle, draw, minimum size=0.3cm, inner sep=1}]
    \node[state, shape=ellipse] (qOmc)               {$q \Omega c$};
    \node[state, shape=ellipse] (qDc0) [right of=qOmc, shift={(0.5, 0)}] {$qDc0$};
    \node[state, shape=ellipse] (qRc100) [right of=qDc0, shift={(-0.5, 0)}] {$qRc100$};

    \draw (qOmc) edge [above]node[align=center, shift={(0, -0.5)}]{$\mathit{select}(b)/$\\$Beverage(b,t:=0,p:=100)$}  (qDc0);
    \draw (qDc0) edge [reflexive above, out=170, in=140]node[align=center, shift={(-0.7, 0)}]{$\mathit{coin}(c)[c > p]/Reject(c)$} (qDc0);
    \draw (qDc0) edge [reflexive above, out=10, in=40]node[align=center, shift={(0.1, 0)}]{$\mathit{vend}()/\omega$} (qDc0);
    \draw (qDc0) edge [above] node[align=left, shift={(0, -0.5)}]{$\mathit{coin}(c)[c \le p]/$\\$Display(t:=c)$} (qRc100);
    \draw (qRc100) edge [reflexive above, out=170, in=140]node[align=left, shift={(0, 0)}]{$\mathit{coin}(c)[t+c > p]/$\\$Reject(c)$} (qRc100);
    \draw (qRc100) edge [reflexive above, out=10, in=40]node[align=center, shift={(-1, 0)}]{$\mathit{vend}()[t \le 50]/\omega$} (qRc100);
    \draw (qRc100) edge [reflexive below]node[align=center, shift={(-0.1, 0.0)}]{$\mathit{coin}(c)[t+c \le p]/$\\$Display(t:=t+c)$} (qRc100);
    \draw (qRc100) .. controls ($(qRc100)+(200:3.5)$) and ($(qOmc)+(320:1.3)$) .. (qOmc) node[below, pos=0.4]{$\mathit{vend}()[t > 50]/Serve(b,t:=0,p:=0)$};
\end{tikzpicture}
    \caption{The generalised inferred EFSM}
  \negativespace
    \label{fig:drinks-inferred}
\end{figure}
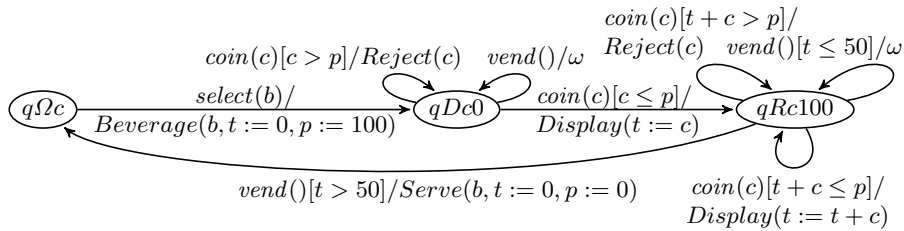

At this point, the sampled FSM is complete, so the \ehW{} global algorithm calls the \textsc{generalise} procedure that uses genetic programming to generalise from the collected samples to get the EFSM, as was done in \cite{Foster2023ICFEM}. The algorithm stops with the final conjectured EFSM shown in Fig.~\ref{fig:drinks-inferred}.

This is not exactly the same as the machine shown in Fig.~\ref{fig:runningexample}, but it is behaviourally equivalent for our domain of available experiments. The main difference lies in the number of control states. Interestingly, the 3-state model seems more ``natural'' for a human eye, because it distinguishes the situation where we have just selected a drink, from the situation where we have already started paying for the drink.

\section{Conclusion}\label{sec:conclusion}
In this paper, we investigated how to infer EFSM models that include registers in guards. By allowing registers to be used in guards, the inference method presented here should be applicable to many systems. It is less restrictive than our previously proposed method~\cite{Foster2023ICFEM}, but we add the assumption of register observability.

The benefit from inferring EFSM models vs using existing algorithms that infer simple automata (DFA) or FSM is to avoid state explosion that would result from expanding the input alphabet to all possible values of parameters, which would not be applicable to infinite domains anyway. Moreover, we can infer expressions for guards and functions, whereas a flattened space resulting from expanding the alphabet would only provide pointwise functions and relations.

\small
% \fxnote{The commands below removes the space between the refs. the 6pt can be adjusted; 0pt means the original space.}
\let\oldbibitem=\bibitem
\def\bibitem{\vspace*{2pt}\oldbibitem}
\bibliography{bibliography}

\newpage
\appendix
\renewcommand{\thesection}{\appendixname~\Alph{section}}

\section{Algorithms}\label{app:algorithms}

\setcounter{AlgoLine}{0}
\newcommand{\SUL}{\textit{SUL}}
\newcommand{\CEx}{\textit{CE}}

% \SetKwFunction{ehW}{\textrm{\textsc{ehW}}}

% Simplified ehW algorithm
\begin{algorithm}[hbt!]   
    \SetAlgoLined
    \nl \KwIn{%
            $I_1 \subset I_s \subset \mathcal{I}$,
            $W \subset I_1^*$,
            $h \in I_1^*$} 
            \Comment{$h$ and $W$ can be empty, or hints could be provided} \\
    \nl \Init{$T \gets \epsilon$} 
% Actually R_g, R_w are not used in the algo, so we do not add the following
% , $R_w = R_g = R$
% Note however that rho_g and rho_w are used in Backbone and Transfer,
% but this can be explained elsewhere.
         \Comment{T is the learning trace}\\
    \nl  \label{salg:repeat}
         \Repeat{no counterexample found}{%
    \nl      $Q, \Delta, \Lambda \gets \emptyset$ \\
    \nl      \Repeat{\textsc{Backbone} terminates with no inconsistency}{%
    \nl         $T,Q,\Delta,\Lambda \gets$ \textsc{Backbone} ($T,I_1,I_s,h,W,Q,\Delta,\Lambda$) \\
                % With assumption h and W are correct, no update needed below
    \nl         handle inconsistencies on the way to update $h,W,I_1$ 
                \Comment{if h \& W not trustable} \\
    \nl         $(T,\CEx) \gets \textsc{GetNFSMCounterExample}(T,Q,\Delta,\Lambda,\SUL,I_s)$ \\
                Ask for a \CEx, preferably prioritizing inputs from $I_1$ \\
    \nl         \If{\rm{\CEx\ found}}{%
    \nl              $(W,I_1,I_s,Q,\Delta,\Lambda) \gets$ \textsc{ProcessCounterexample} \\
                    \Comment{If W changed, this resets $Q,\Delta,\Lambda$} \\
                    \Comment{At this point, $I_s$ would not be changed} \\
    \nl             continue to start of repeat \textsc{Backbone} loop
                 }
              }
% Still wondering whether we should reduce after each backbone or just before generalizing.
% It makes sense to try finding CE on a smaller structure, but as there is no unique reduction, could this trigger instability ?
% At the same time, sampling on a non-reduced NFSM provides more diverse samples, even
% if some are redundant once reduced. So we do it just for helping generalise.
% For sure, following our experiment with Michael's tool, it is definitely useful to
% reduce before calling generalise.
    \nl     $(Q,\Delta,\Lambda) \gets \textsc{ReduceFSM} (Q,\Delta,\Lambda)$ \\
    \nl     \Repeat{$\neg$ (\CEx\ is a data \CEx)}{%
    \nl         $M \gets$ \textsc{generalise}($T,h,Q,I,O,P_I,P_O,\Delta,\Lambda$) \\
    \nl         $(T,\CEx) \gets \textsc{GetCounterExample}(M,\SUL)$ 
            }
    \nl     \If{\rm{\CEx\ found}}{%
    \nl         $(W,I_1,I_s,Q,\Delta,\Lambda) \gets$ \textsc{ProcessCounterexample}
            \Comment{If W modified, $Q,\Delta,\Lambda$ are reset}
           } 
        }
    \nl     \Return{$M$}
~ \\
~ \\
    \caption{Global \textit{ehW} algorithm}
    \label{alg:full-algo}
\end{algorithm} % Simplified ehW algorithm

\setcounter{AlgoLine}{0}

\begin{algorithm}[hbt!] 
   \nl \Fn{\textsc{Backbone}($T,I_1,I_M,h,W,Q,\Delta,\Lambda$)}{%
   \nl    \Init{$H \gets \emptyset$, $J \gets I_1$,
             $q \gets \bot$, $r \gets \bot$} \\
%   \nl \Repeat {$\Delta,\Lambda$ are complete over $I_M$ on an Scc 
    \nl    \Repeat{no further sampling reachable in current Scc
          \label{backbone:repeat}}{%
   \nl    \If{$q = \bot$}{%
             \Comment{We do not know where we are} \\
   \nl       $(T,q,r) \gets \textsc{Home}(T,r,h)$
          }
   \nl $(q',r',X,Y,r_1) \gets \textsc{Transfer}(q,r,\Delta,\Lambda,I_1,J\setminus I_1)$ \\
       \Comment{Target next transition to learn/sample} \\
   \nl \If{such a path cannot be found}{%
   \nl    $J \gets I_M$ 
          \Comment{Graph not connected, try connecting with $I_M$}} 
   \nl \Else{
   \nl    \If{$Y = \Omega$ or $\omega$}{\Comment{$\pi(X)$ is not enabled in $q',r'$} \\
   \nl       $\Lambda(q',\pi(X),\omega) \gets \{r',X,Y,r_1\}$, $\Delta(q',\pi(X),Y) \gets q'$ \label{backbone:Omega} \\
   \nl       $q \gets q'$, $r \gets r'$ \Comment {We continue learning from the same state}
          }
   \nl       \Else{
                \Comment{$ {(q',r')} - X/Y \to \mathbf{(q'',r_1)} \to w/\xi \to (q''',r_2)$} \\
   \nl          Let $q'' = \Delta(q',\pi(X),\pi(Y))$ \\
                \Comment{$q''$ (and so $\Delta$) might be a ``state under construction''} \\
   \nl          \If{$q''$ is fully defined}{%
                    %and $\exists X',Y'$ s.t. $\pi(Y)=\pi(Y')$ and $\pi_2(\Lambda(q',X'))=Y'$}
                   \Comment{We are sampling new values of a known transition with same abstract output} \\
   \nl             $\Lambda(q',\pi(X),\pi(Y)) \gets \Lambda(q',\pi(X),\pi(Y)) \cup \{(r',X,Y,r_1)\}$,
                   \Comment{Note we should not have a different $Y'$ with same $r'$ unless (W-)inconsistency} \\
   \nl             $q \gets \Delta(q',\pi(X),\pi(Y))$, $r \gets r_1$
                }
   \nl          \Else{
                   \Comment{Learn tail of transition from q' on input X} \\
   \nl             for some $w \not\in dom (\pi_1(q''))$, 
                   \label{backbone:learn-trans} \\
   \nl             \qquad $(T,\xi,r_2) \gets \textsc{Apply}(T,r_1,w)$
                   \label{backbone:add-to-dictionary} \\
   \nl             $\Lambda(q',\pi(X),\pi(Y)) \gets \Lambda(q',\pi(X),\pi(Y)) \cup \{(r',X,Y,r_1)\}$ \\
   \nl             $\pi_1(q'')(w)) \gets \pi(\xi)$ \label{backbone:update-delta} 
                   \Comment{This updates $\Delta$}  \\
   \nl             \If{$dom(\pi_1(q''))=W$}{%
                      \Comment{If $R_W=R$, $\rho_w(r,\epsilon)=r$ is simply register configuration, i.e. tuple of register values.} \\
   \nl                $Q \gets Q \cup \{(\pi_1(q''),\rho_w(r_1,\epsilon)\}$ \\
   \nl                $A \gets \textsc{UpdateAccess}(T,q'',\rho_w(r_1,\epsilon))$ \\
                      \Comment{We record the shortest access to $(q'',r_1)$ from $(q,r)$ or the last homing in $T$} \\
                      % Note that Delta(q',X.w) could be undefined so q would also get \bot
   \nl                $q \gets \Delta^*(q',\pi(X.w),\pi(Y.\xi))$ if defined else $q \gets \bot $ \\
   \nl                $r \gets r_2$ 
                    }
   \nl              \Else{
   \nl                 $q \gets \bot$
                    }
                }
   \nl      \If {$\Delta^-(\Delta,\Lambda,h,a,A)$ (where $h/a$ was latest homing in $T$) is defined and contains a complete strongly connected component (Scc)}{%
% We let the calling algorithm (ehW) decide whether we jump to full sampling space
% at once, or progressively adding new values to sample.
   \nl          $J \gets I_M$ \\
                \Comment{Learning in Scc more samples and possibly new transitions with an increased set of inputs}
                \label{backbone:Scc}
             }

            }
        }
   }
   \nl \Return $T,Q,\Delta,\Lambda$
   }
~ \\
~ \\
   \caption{Backbone procedure with guards on registers}
   \label{alg:backbone}
\end{algorithm} % backbone

\setcounter{AlgoLine}{0}

\SetKwFunction{Transfer}{\textrm{\textsc{Transfer}}}

\begin{algorithm}[hbt!] 
   \nl \Fn{\Transfer($q, r, \Delta, \Lambda, I_1, I_b$)}{%
          \Comment{We look for deterministic transfer, either through 
                   unguarded transitions, or when registers and input 
                   determine known transitions} \\
   \nl    \Comment{To reach a given input configuration $r'$, we record 
                   the list of parameter values that can be set by unguarded 
                   transitions while building the path, and can adapt the 
                   parameter values at the end to set values to $r'$} \\
   \nl     \Comment{First we look for a short transfer with a bounded search 
                    ($k \geq 0$ is the bound, tailorable), and if that fails, 
                    we resort to a path from re-homing}\\
% For easier implementation, we could just start implementing with k=0.
% Actually alpha is a bit special because it contains concrete inputs but only abstract outputs. We may are not fully consistent here?
           \label{transfer:shortest-alpha}
   \nl     Find short(-est) $\alpha \in (\mathcal{I} \times O)^k$ and 
           $X \in I_1$ with $\Delta^*(q,\pi(\alpha))=q'$ and 
           $\rho_w(r,\alpha)=\pi_2(q')=\rho_w(r',\epsilon)$ 
           s.t. $\pi_1(\Delta(q',\pi(X),*))$ is partial, \\
           \Comment{First try to learn new input from a state, in its 
                    reference configuration}\\
   \nl     \Comment{``partial'' means either not defined at all, or there 
                    is an output whose characterisation is partial; 
                    if $\Omega$ was the output for $\pi(X)$ it is not partial 
                    regardless of $r'$ and $X$; if it was $\omega$ for a 
                    given $r'$ and $X$, we can still look for a different 
                    $r'$ or $X$.} \\
   \nl     or $\exists r' \neq \pi_2(q'), Y \neq Y'$
           s.t. $\rho_g(r,\alpha)=\rho_g(r',\epsilon)$
           and $\{(\pi_2(q'),X,Y,*),(r',X,Y',*)\} \subset 
              \Lambda(q',\pi(X),\pi(Y))$
           and $\pi_1(\Delta(q',\pi(X),\pi(Y')))$ is partial 
           \Comment{or a guarded transition} \\
   \nl        \If{previous fails}{%
                 \Comment{(otherwise) no transition to learn in current Scc} \\
   \nl           Find $\alpha$ and $X \in I_b$ s.t. 
                    $\Lambda(q',\pi(X),*)$ does not contain $(*,X,*,*)$ \\
                 \Comment{transfer to a transition to be sampled} \\
   \nl           $q'=\Delta^*(q,\pi(\alpha))$, $r'=\rho(r,\alpha)$
              } 
   \nl        \If{all previous fails}{%
                 \Comment{bounded search failed, resort to homing} \\
   \nl           $(T,a,r) \gets \textsc{Apply}(T,r,h)$ and 
                 update $\Lambda$ on the way (if transitions are in $Q$) \\
   \nl           \If{$\Delta^-(\Delta,\Lambda,h,\pi(a),A)$ no longer contains 
                    unsampled states and transitions}{%
   \nl               \Return{no path found} 
                     \Comment{all transitions in reachable Scc already 
                              sampled on $I_b$, transfer fails}
                 } 
   \nl           Look by BFS for shortest sequence $\alpha$ in 
                    $\Delta^-(\Delta,\Lambda,h,\pi(a),A)$, 
                    pick $A(\pi(a))(q',r')=\alpha$ and $X$
                    s.t. $\pi_1(\Delta(q',\pi(X),*))$ is partial or has 
                    guarded transition to partial state, or failing that has 
                    a transition to be sampled
             }
             \Comment{Here $q',r',\alpha, X$ are defined. If $\Delta$ was 
                      partial in $q$, then $\alpha=\epsilon$, $q'=q$} \\
% We are in q,r; alpha will lead us to q',r' where we need to learn transition on X
% and by X/Y we will go to q'',r1.
% We overload Apply to accept not just input sequences, but IO sequences
% Now we attempt transferring
   \nl       $(T,\beta,r') \gets \textsc{Apply}(T,r,\alpha)$ and 
                update $\Lambda$ on the way \\
%            \Comment{State is now $q'$ unless some abstract output differed}
   \nl       \If{$\beta \neq \pi_o(\alpha)$}{%
                \Comment{Transfer stopped prematurely on differing output} \\
   \nl          let $\beta=\beta'.o'$, 
                $\alpha=\alpha'.X.\alpha''/\beta'.o.\beta''$ s.t. 
                $\pi_o(\alpha')=\beta', o' \neq o$ \\
   \nl          $Y \gets o'$; $r_1 \gets r', 
                r' \gets \rho(r,\alpha'), 
                q' \gets \Delta^*(q,\alpha')$ \\
% IMPORTANT note: thanks to observability and W being assumed correct, we know the LAST transition was guarded
% If W was not correct, then it could be a W-ND and could come from an earlier transition
                \Comment{We found a new guarded transition} \\
   \nl          \Comment{If h or W were not correct, we should handle 
                         inconsistency if after $\alpha'$ we have same 
                         $R_g$ configuration $\rho_g(r)$ for $o$ and $o'$}
            }
   \nl      \Else{%
               \Comment{$(q,r) - \alpha(')/\beta(') \to 
               \bold{(q',r')} - X/Y \to (q'',r_1) \to w/\xi \to (q''',r_2)$} \\
   \nl         $(T,Y,r_1) \gets \textsc{Apply}(T,r',X)$ \\
            } 
   \nl      $A \gets \textsc{UpdateAccess}(T,q',r')$ 
            \Comment{We record the shortest access to $(q',r')$ from 
                     $(q,r)$ or the last homing in $T$} \\
   \nl      \Return $(q',r',X,Y,r_1)$
    }
~ \\
~ \\
    \caption{Transferring from $q,r$ to next transition to learn}
    \label{func:transfer}
\end{algorithm}

\FloatBarrier

\section{Drinks full inference trace}\label{app:drinks_inference}
\begin{center}

\begin{tabular}[|c|]{l}
	$
\state[s0]{q}
\X[X=coin(100)]{
\event{\Omega}{1}
}
\state[s1]{q}
\X[X=select(\textit{coffee})]{
\event{Beverage(\textit{coffee},0,100)}{2}
}
\state[s1]{q}
\X[X=vend]{
\event{\omega}{3}
}
\state[s1]{?}
\step{s=coin(50)}{
\event{Display(50)}{4}
}

	$
\end{tabular}
\end{center}

The algorithm starts by homing (here no input to apply as $h$ is empty), then characterising the state reached. Since the characterisation set is empty, we are in the only possible state, which we name $q$. From $q$, we learn the (abstract) output associated to each (concrete) input from $I_1$, coming up with the initial ``daisy'' conjecture at Step~3. A so called ``daisy'' machine is a single state machine with as many transitions as inputs, all looping back on the state.

To learn the correct output function, we now sample with $I_s$. Trying the first element in $I_s$, viz. $coin(50)$ (Step~4) yields output $Display$ instead of the abstract output $\Omega$ that had been learnt for the transition on that input in the daisy. This is a typical case of \WND{}, i.e. a counterexample provided for free by sampling, without having to call the oracle. When processing this counterexample, we first investigate whether the different output could be due to a different configuration of registers, but this cannot be the case with $\Omega$, i.e. we are in a control state that cannot accept the input $coin$ (regardless of input parameters and registers). This means that $coin$ separates at least two states, so we restart the algorithm with $W={coin(100)}$, favouring the concrete input of $coin$ that is already in $I_1$. And since this is also a case of \hND{}, we also update $h=coin(100)$.
We now restart the backbone algorithm with these new values of $h$ and $W$.

\begin{tabular}[c]{l}
  $
\state[s1]{?}
\h[h=coin(100)]{
\event{Reject(100)}{5}
}
\state[s1]{?}
\W[W=coin(100)]{
\event{Reject(100)}{6}
}
\state[s1]{?}
\h[h=coin(100)]{
\event{Reject(100)}{7}
}
\state[s1]{qRc50}
\X[X=coin(100)]{
\event{Reject(100)}{8}
}
\state[s1]{?}
\W[W=coin(100)]{
\event{Reject(100)}{9}
}
  $
\end{tabular}

In Steps~5--6, we learn a homing tail: $H(Reject,(\textit{coffee},100,50,100))$ is defined as the pair $(Reject,(\textit{coffee},100,50,100))$, i.e. after seeing the abstract output sequence $Reject$ the configuration (state and registers) we reach is characterised by the abstract output sequence $Reject$ when $W$ is executed in the register configuration $(\textit{coffee},100,50,100)$, and the register configuration is unchanged (because the output $Reject(100)$ does not change the value of the $c$ register shared with the input $coin(100)$). For convenience, we name this state $qRc50=(Reject,(\textit{coffee},100,50,100))$ for short. Note that this is a state of the conjecture, but that conjecture state also encodes the register configuration that was reached when we learnt that response to the characterization.

After applying $W$ at Step~6, we do not know what is the current state, so we start by homing again, and it turns out that the response to the homing sequence is the same that we already learnt, so we know that after Step~7, we are again in the learnt state that we called $qRc50$.
Now we can start learning a transition from that state, and we send at Step~8 the first input from $I_1$, viz. $coin(100)$, immediately followed by application of $W$. As we get again the response $Reject(100)$ to $W$ AND we reach the same configuration of registers $(\textit{coffee}, 100,50,100)$, we learn that the tail state of the transition is the same as the initial state $qRc50$. So we have learnt a looping abstract transition $coin/Reject$ on state $qRc50$ and we also collected a sample $coin(100)/Reject(100)$ for that transition with initial and final register configurations $(\textit{coffee}, 100,50,100)$.
\footnote{Since that transition is looping, we might know what is the current state in the conjecture, so we would not need to home again. But our algorithm currently does not implement this optimization.}

\begin{tabular}[c]{l}
  $
\state[s1]{qRc50}
\h[h=coin(100)]{
\event{Reject(100)}{10}
}
\state[s1]{qRc50}
\X[X=select(\textit{coffee})]{
\event{\Omega}{11}
}
\state[s1]{qRc50}
\X[X=vend]{
\event{\omega}{12}
}
\state[s1]{?}
\s[s=coin(50)]{
\event{Display(100)}{13}
}
\state[s1]{?}
\W[W=coin(100)]{
\event{Reject(100)}{14}
}
  $
\end{tabular}

At Step~10, we home again, and we know we are still in the same conjecture state $qRc50$, so we can continue learning transitions for other abstract inputs from $I_1$. At Step~10 we apply $select(\textit{coffee})$, and since this is an $\Omega$-transition, we know that in state $qRc50$ input $select$ is not accepted and the configuration of the SUL is unchanged, so we can directly apply the next input from $I_1$, $vend$ that is not accepted either.
At this point, we have learnt a new daisy machine, this time from state $qRc50$, with a single non-$\Omega$ transition, $coin/Reject$. This is a strongly connected component, so we start sampling with a new concrete value $coin(50)$ and now we observe a different output (Step~13). The algorithm prioritizes considering that this is a guarded transition $coin/Display$ from $qRc50$ rather than an error on the state that could have been caused by insufficient characterization, so we first try to learn the tail state of that guarded transition by applying $W$. At Step~14 we observe the same output $Reject$ for $W$, but now the register configuration reached is $(\textit{coffee},100,100,100)$ so this is a new conjecture state that we name $qRc100=(Reject,(\textit{coffee},100,100,100))$. Actually, both states $qRc50$ and $qRc100$ of the conjecture map to the same state $s_1$ of the SUL, but since register configurations are different and they might have influenced the response to $W$, the algorithm conservatively ``duplicates'' the states: we can only compare characterizations applied from the same register configuration.

\begin{tabular}[c]{l}
  $
\state[s1]{?}
\h[h=coin(100)]{
\event{Reject(100)}{15}
}
\state[s1]{?}
\W[W=coin(100)]{
\event{Reject(100)}{16}
}
\state[s1]{?}
\h[h=coin(100)]{
\event{Reject(100)}{17}
}
\state[s1]{qRc100}
\X[X=coin(100)]{
\event{Reject(100)}{18}
}
\state[s1]{?}
\W[W=coin(100)]{
\event{Reject(100)}{19}
}
  $
\end{tabular}

When homing again at Step~15, we find a new response to the homing sequence, so we identify the tail configuration at Step~16, and we learn the tail was $H(Reject,(\textit{coffee},100,100,100))=qRc100$. Reapplying $h$, we reach this time $qRc100$, from which we learn the transition $coin/Reject$.

\begin{tabular}[c]{l}
  $
\state[s1]{qRc100}
\X[X=select(\textit{coffee})]{
\event{\Omega}{20}
}
\state[s1]{qRc100}
\X[X=vend]{
\event{Serv(\textit{coffee},0,0)}{21}
}
\state[s0]{?}
\W[W=coin(100)]{
\event{\Omega}{22}
}
\state[s0]{q \Omega c}
\X[X=coin(100)]{
\event{\Omega}{23}
}
$\\$
\state[s0]{q \Omega c}
\X[X=select(\textit{coffee})]{
\event{Beverage(\textit{coffee},0,100)}{24}
}
\state[s1]{?}
\W[W=coin(100)]{
\event{Display(100)}{25}
}
\state[s1]{?}
\h[h=coin(100)]{
\event{Reject(100)}{26}
}
\state[s1]{qRc100}
\transfer[transfer=vend]{
\event{Serv(\textit{coffee},0,0)}{27}
}
  $
\end{tabular}

Since the last transition is looping, the state after Step~19 is still $qRc100$, from which we learn transitions on the other two inputs. After $vend/Serve$ we have reached a new state $q \Omega c$ from which $coin$ is not accepted. Although Step~23 could be avoided with an optimization, the backbone algorithm without optimization will relearn the $\Omega$ transition for $coin$ in that state. From $q \Omega c$, we learn the $select/Beverage$ transition (Step~24) that leads to a new state that is identified after Step~25 as $qDc0=(Display,(\textit{coffee},100,0,100))$. After homing again, we know we reach state $qRc100$ after Step~26. Since we already learnt all 3 transitions from that state, we transfer by $vend/Serve$ to the nearest state $q \Omega c$ from which we still need to learn one transition, on input $vend$.

\begin{tabular}[c]{l}
  $
\state[s0]{q \Omega c}
\X[X=vend]{
\event{\Omega}{28}
}
\state[s0]{q \Omega c}
\transfer[transfer=select(\textit{coffee})]{
\event{Beverage(\textit{coffee},0,100)}{29}
}
\state[s1]{qDc0}
\X[X=coin(100)]{
\event{Display(100)}{30}
}
\state[s1]{?}
\W[W=coin(100)]{
\event{Reject(100)}{31}
}
$\\$
\state[s1]{qRc100}
\transfer[transfer=vend]{
\event{Serv(\textit{coffee},0,0)}{32}
}
\state[s0]{q \Omega c}
\transfer[transfer=select(\textit{coffee})]{
\event{Beverage(\textit{coffee},0,100)}{33}
}
\state[s1]{qDc0}
\X[X=select(\textit{coffee})]{
\event{\Omega}{34}
}
  $
\end{tabular}

Since $vend$ is not accepted in state $q \Omega c$, we stay there and must transfer to the remaining incomplete state $qDc0$ that was first reached after Step~24. After learning $coin/Display$ from there, that reaches state $qRc100$ (after Step~30), we know that Step~31 was a looping transition (learnt at Steps~18-19), so now we need again to transfer to the incomplete state $qDc0$, and the shortest transfer in the partial conjecture graph is a sequence $vend.select$ (Steps~32-33).

\begin{tabular}[c]{l}
  $
\state[s1]{qDc0}
\X[X=vend]{
\event{\omega}{35}
}
\state[s1]{qDc0}
\s[s=coin(50)]{
\event{Display(50)}{36}
}
\state[s1]{qRc100}
\s[s=coin(50)]{
\event{Display(100)}{37}
}
\state[s1]{?}
\W[W=coin(100)]{
\event{Reject(100)}{38}
}
\state[s1]{qRc100}
$\\$
\state[s1]{qRc100}
\transfer[transfer=vend]{
\event{Serv(\textit{coffee},0,0)}{39}
}
\state[s0]{q \Omega c}
\s[s=select(tea)]{
\event{Beverage(tea,0,100)}{40}
}
\state[s1]{qDc0}
\step{CE=coin(50).vend}{
\event{Display(50)}{41}
\event{\omega}{42}
}
  $
\end{tabular}

At Step~34 we learn that input $Select$ is not accepted from $qDc0$, and at Step~35 that $vend$ is not accepted at least in that register configuration. We have now transitions for all inputs from all 3 states $q \Omega c$, $qDc0$ and $qRc100$. We now just need to sample the non-$\Omega$ transitions with values from $I_s$. The last homing brought us into state $qRc100$ and the strongly connected component (Scc) has 3 states (it no longer includes $qRc50$). The backbone algorithm terminates at Step~40 with a fully sampled Scc, so the main ehW algorithm asks first for a counterexample on the control NFSM machine.

At Step~42, we discover a guarded transition from state $qRc100$: $vend/\omega$ (whereas previously we had $vend/Serve$ from that state). Since input $vend$ has no parameter, the guard can only be a condition on registers. For $vend/Serve$ we had register configuration $(\textit{coffee},100,100,100)$ (in Steps~21, 27, 32, 39), whereas we get an $\omega$ output from register configuration $(tea, 50,50,100)$.

The algorithm has easily handled the counterexample since it is in a self loop, so it just adds this transition.
At this point, the inferred NFSM still misses a guarded transition on $qDc0$, for which we need a coin with a value greater than 100. The \textsc{GetCounterExample} procedure provides the following NFSM counterexample.

\begin{tabular}[c]{l}
  $
\state[s1]{qRc100}
\step{CE=coin(50).vend.select(\textit{coffee}).coin(200)}{
\event{Display(100)}{43}
\event{Serv(tea,0,0)}{44}
\event{Beverage(\textit{coffee},0,100)}{45}
\event{Reject(200)}{46}
}
\state[s1]{?}
\W[W=coin(100)]{
\event{Display(100)}{47}
}
  $
\end{tabular}

After learning the tail state of the guarded transition $coin(200)/Reject(200)$ from state $qDc0$, we see it is a loop transition.

\end{document}